\documentclass[aps,prd,showpacs,nofootinbib,floats,floatfix,preprintnumbers,groupedaddress,twocolumn]{revtex4}

\usepackage{bm}
\usepackage{latexsym}
\usepackage{dcolumn}
\usepackage{amsmath,amsfonts,amssymb}
\usepackage{graphicx,epsfig}
\usepackage{amsmath}
\usepackage{fancyhdr}
\usepackage{hyperref}
\usepackage{graphicx,epstopdf}
\hypersetup{
	colorlinks   = true, 
	urlcolor     = blue, 
	linkcolor    = blue, 
	citecolor   = red 
}

\def\o{\omega}

\def\no{\nonumber}
\def\a{\alpha}
\def\b{\beta}

\def\d{\delta}

\def\p{\partial}

\def\na{\nabla}

\def\lie{\pounds_{\xi}}
\def\t{\tilde}

\def\R{\mathcal{R}}
\def\ie{{\it i.e.~}}
\def\etc{{\it etc.}}
\def\dif{diffeomorphism~}
\def\mg{\sqrt{-g}~}
\def\adt{$(\d E^{ab})\xi_b$~}
\def\la{Lagrangian~}
\def\kv{Killing vector}
\def\ll{Lanczos-Lovelock~}
\def\ckv{conformal Killing vector}

\def\J{\mathcal{J}}
\def\G{\Gamma}
\def\K{\mathcal{K}}
\def\H{\mathcal{H}}

\begin{document}
\title{Abbott-Deser-Tekin like conserved quantities in Lanczos-Lovelock gravity: beyond Killing diffeomorphisms}
\author{Krishnakanta Bhattacharya\footnote{\color{blue} krishnakanta@iitg.ac.in}}
\author{Bibhas Ranjan Majhi\footnote {\color{blue} bibhas.majhi@iitg.ac.in}}

\affiliation{Department of Physics, Indian Institute of Technology Guwahati, Guwahati 781039, Assam, India}

\date{\today}

\begin{abstract}
We obtain the conserved Abbott-Deser-Tekin (ADT)-{\it like} current for the Lanczos-Lovelock gravity for a diffeomorphism vector, which defines the horizon of a spacetime and, importantly, is not necessarily a Killing vector. As the original ADT current is defined only for the presence of a background Killing vector, one cannot use it extensively for the thermodynamic description of the wide classes of non-Killing horizons which appear in gravity. On the other hand, this general approach can be utilized for those horizons. Here, the conserved current can be written as the derivative of the two-rank anti-symmetric potential, the connection of which is apparent with the conserved Noether potential from our analysis. If one assumes the diffeomorphism vector as the Killing one, the results match to the ADT case, whereas non-trivial result comes for the conformal Killing vectors and other horizon defining \dif vectors. 
\end{abstract}


\maketitle

\section{Introduction}
The conserved charges \cite{Komar:1958wp, Arnowitt:1962hi, Wald:1993nt, Abbott:1981ff, Deser:2002rt, Deser:2002jk, Deser:2003vh} have played a crucial role in the theory of black hole since the time when the black holes are identified as the thermodynamic objects with  having proper thermodynamic descriptions \cite{Bekenstein:1973ur, Hawking:1974sw, Bardeen:1973gs}. In the theory of the black hole mechanics, the conserved charges often provides the physical thermodynamic quantities, such as the entropy, energy, angular momentum \etc~There are two major approaches to develop the thermodynamic description of a black hole from the conserved charges. Among them, the conserved Noether current due to the \dif~has enjoyed the central attention over the years. Following the Wald's \cite{Wald:1993nt} prescription, one can successfully obtain the first law of the black hole thermodynamics. In this approach, the entropy is defined by the conserved Noether charge on the black hole horizon, whereas the mass and the angular momentum is defined on the asymptotic infinity by the same conserved current along with a correction factor.

 There is another elegant approach developed independently in General theory of Relativity (GR) for the formulation of the thermodynamic description from the conserved charges, which is popularly known as the Abbott-Deser-Tekin (abbreviated as ADT) approach \cite{Abbott:1981ff, Deser:2002rt, Deser:2002jk, Deser:2003vh, Deser:2006gt}. This formulation works well not only for asymptotically flat black hole spacetimes in GR, but also for the anti-deSitter (AdS) spacetimes in GR \cite{Abbott:1981ff} as well as higher order gravity theories \cite{Deser:2002rt, Deser:2002jk, Deser:2003vh} (For the thermodynamic description using the ADT approach, also see the recent works \cite{Bouchareb:2007yx, Kim:2013cor, Kim:2013zha, Hyun:2014kfa, Gim:2014nba, Hyun:2014sha, Hyun:2014nma, Hyun:2016dvt, Hyun:2016isn, Hyun:2017nkb, Bhattacharya:2018xlq}). Recently, the ADT approach has been extended successfully by us in the scalar-tensor theory as well, where we have also shown that the two approaches (i.e. Noether and ADT formalisms) are indeed equivalent \cite{Bhattacharya:2018xlq}. In the latter approach, one formulates a conserved current accounting the variation of the equation of motion due to the variation of the metric tensor. Remember, till now both the formulation are suitable when there is a Killing symmetry of the underlying spacetime. Apart from this two approach (Wald and ADT), in \cite{Barnich:2001jy} another method has been described, which belongs to the Wald's class, but with some different boundary conditions. Here, the asymptotic symmetries and conservation laws has been investigated using covariant phase space method.
 
 Before discussing the main aim of the present paper, we want to mention couple of important facts which has been observed in exploring the different features of the gravitational paradigm. This will give the motivation of the present work. Soon after the observation of the thermodynamic nature of the black hole horizon, Unruh showed that an accelerated observer can find radiation in the Minkowski vacuum \cite{Unruh:1976db} and consequently this observer is able to associate temperature and entropy on the Rindler horizon (obtaining the entropy of the Rindler horizon using Virasoro algebra has shown recently in \cite{Majhi:2012tf, Majhi:2017fua}). Moreover, assigning of the first law of thermodynamics on it leads to Einstein's equations of motion \cite{Jacobson:1995ab}. Later investigation revels that such feature not only restricted to a Rindler kind of horizon, any generic null surface also incorporates thermodynamic structure (see \cite{Padmanabhan:2010xe, Parattu:2013gwa, Chakraborty:2015aja, Chakraborty:2016dwb, Bhattacharya:2018epn} for recent developments in this direction). Note that in the generic null case, the surface defining vector (here it is a null vector) may not be, in general, a Killing one.

All these works suggest that the thermodynamic description of a spacetime is not restricted to the Killing horizon. Instead, the thermodynamic structure is an extensive feature in the horizon thermodynamics, which is true irrespective of any specific horizon (including the Killing horizon and other non-Killing horizon such as apparent horizon, trapped horizon \cite{Nielsen:2008cr}, etc.  present in GR). However, the existing approach of defining the ADT current fails to be befitting in the thermodynamics of the wide classes of non-Killing horizon.  Remember, in literature, the conserved Noether current is already defined for Killing and non-Killing \dif vectors (see the project 8.1 of \cite{paddy}). Therefore, the existning Noether approach can be applicable for both Killing and non-Killing horizon surfaces. On the other hand, in the existing formalism, the ADT current is only restricted to the Killing vectors which defines the horizon surface.

 Our aim here is to find the conserved quantities of the gravitational theory for those \dif vectors, the vanishing of norm of which defines those wide classes of non-Killing horizons. We adopt an elegant method to formulate  a quasi-local ADT-like conserved current in \ll gravity \cite{Lanczos:1938sf, Lovelock:1971yv} (often referred to as Lovelock gravity as well) for those horizon defining \dif vectors. The \ll (LL) model of gravity is an extension of Einstein's theory of gravity in higher dimension containing quadratic and higher order polynomials of the curvature tensor. In addition, like GR, the field equations contain only up to the second order derivative of the metric and the theory is free of unphysical ghosts. Moreover, a major motivation of studying the LL gravity comes from the string theory. LL gravity resembles to the string theory inspired gravity models \cite{Zwiebach:1985uq, Boulware:1985wk}. Besides, the quadratic Gauss-Bonnet term is studied in the context of string theory with particular attention given due to its ghost free solution \cite{Boulware:1985wk}. In summary, the LL gravity represents the higher order generalization to the Einstein's gravity which is studied to examine how the effect of gravity gets modified at short distance in presence of higher order curvature terms in the action. As a result, the LL gravity is studied in several contexts of gravity including in the context of thermodynamics as well (see the review \cite{Padmanabhan:2013xyr}). 

 The process here we follow (to obtain the ADT-like current) is different from the original ADT method and will show that our current and anti-symmetric potential (we call as the ADT-like potential) both reduce to the usual conserved ADT current and potential respectively, when we take the horizon defining \dif vector as the Killing one. In addition, we shall find the explicit relation between the  Noether and present ADT-like anti-symmetric potentials. This relation will help to understand more about those potentials. The whole analysis is very general as the entire approached is made for the LL gravity. Of course, one can easily obtain the corresponding results for the GR or in Gauss-Bonnet gravity from our analysis by taking the proper limit. Moreover, the \dif vector can be chosen according to the type of the horizon surface.
 
 Finally, we shall discuss two situations: one is the spacetime has Killing symmetry and the other one corresponds to conformal Killing symmetry of the spacetime. In that case one needs to choose the diffeomorphism vector as \kv~(KV) and a \ckv~(CKV), respectively. Interestingly, we show that both for \kv~ and \ckv,~the conserved anti-symmetric  potential is related to the Noether counter part in exactly same manner. However, we shall notice that the expression of the current is not the same for the both cases.
 
 In the following, we start our analysis. Firstly, we obtain the general expression of the conserved current and the potential for \ll gravity. Then we take two special cases of KVs and CKVs.
\section{Conserved quantities for a horizon defining \dif  vector}
It is often proclaimed that any feasible theory of general relativity must be diffeomorphism invariant. Now, since diffeomorphism is an active coordinate transformations, people often tend to refer the diffeomorphism as the ``change in coordinates'', which might be technically correct but, does not imply the complete scenario. The reason is: the operation diffeomorphism implies a set of two consecutive actions--  a pushforward along with a pullback (see appendix B of \cite{carroll} for a detailed discussion). Therefore, it is not an ordinary coordinate transformation. Now it may be noted that most of the theories in physics, which are based on Newtonian gravity or even based on special relativity, does not change its description due to the change in coordinates.  In theses theories the inertial frame enjoys a special status. Whereas, GR or any other gravitational theory, described by the geometry of the spacetime, is build up in such a way that there is ``no preferred set of coordinates''. This implies the theory should be described in a covariant way under any general coordinate transformation, which is usually called as diffeomorphism.

 As the gravity theory has such diffeomorphism symmetry, one usually finds the conserved current and the corresponding charge by the Noether prescription.  It has been observed by Wald \cite{Wald:1993nt} that this conserved current is very useful to obtain the thermodynamic structure of gravity in presence of a Killing horizon. More precisely, the conserved Noether charge, calculated on the Killing horizon for the horizon defining Killing vector, is related to the black hole entropy.  Whereas, this is related to the mass and the angular momentum of a black hole when it is evaluated at asymptotic infinity. All these state that the conserved quantities due to the diffeomorphism symmetry play a crucial role in understanding the thermodynamic description of gravity.

There is another approach developed simultaneously along with the Wald's formalism. This is the ADT formalism which is developed for the presence of a Killing vector in the spacetime. This conserved current is known as the ADT current, which is mentioned in the following.
 Note that when the diffeomorphism vector $\xi^a$ is a Killing one, the term $\J^a=(\d E^{ab})\xi_b$ is a conserved quantity, where 
\begin{equation}
E^{ab}=\R^{ab}-\frac{1}{2}g^{ab}L ~,
\label{EAB}
\end{equation} 
is a second rank tensor in LL gravity, which plays the analogous role to the Einstein tensor in GR. In the above Eq. \eqref{EAB}, $L$ is the Lagrangian of the LL gravity which is a general function of the Riemann tensor $R^a_{~bcd}$ and metric tensor $g^{ab}$ but not of the derivatives of those quantities. Also, $\R^{ab}=P^{acde}R^b_{\,\,\, cde}$ plays the analogous role to the Ricci tensor in GR, where $P^{abcd}=(\frac{\p L}{\p R_{abcd}})_{g_{ij}}$ which satisfies vanishing of covariant derivative; i.e. $\nabla_a P^{bcde}=0$. For more information about $E_{ab}$, consult with the review \cite{Padmanabhan:2013xyr}. In the above expression of $\J^a$, $\d E^{ab}$ represents the linearized tensor, \ie the change of $E^{ab}$ for $g_{ab}\rightarrow g_{ab}+h_{ab}$ up to first order in $h_{ab}$. The conservation of the quantity $\J^a$ ({i.e.}, $\na_a\J^a=0$) follows from the fact that $E^{ab}$ satisfies the general Bianchi identity {\it i.e.} $\na_a E^{ab}=0$ and $\na_a\xi_b$ is an anti-symmetric tensor as $\xi^a$ satisfies the Killing equation. The conserved quantity now popularly known as the ADT current (for more details about ADT quantities, refer to \cite{Abbott:1981ff, Deser:2002rt, Deser:2002jk, Deser:2003vh, Deser:2006gt}).

Both of these currents (the Noether and the ADT) can be utilised for obtaining the thermodynamics of a Killing horizon and, they provide equivalent thermodynamic description (see \cite{Bhattacharya:2018xlq} for an example). But in real scenario, the spacetime is not stationary and consequently, the Killing horizon ceases to exist. In this case one can have other kind of horizons, like apparent horizon, trapped horizon, conformal Killing horizon, generic null surfaces, \etc~None of  the vectors which defines the surface is a time-like Killing vector. Therefore the existing formalism for Noether and ADT does not work in those cases.

As we have mentioned, if $\xi^a$ is not a Killing vector, then \adt is not a conserved quantity anymore. Then above way of defining quantity does not fulfill our aim. Hence we need to adopt a different approach to achieve the goal. Our idea is the following.

First, we identify the Noether current due to the diffeomorphism. Then, we take the variation of the Noether current due to the variation of the metric tensor. Afterwards, we identify the total derivative anti-symmetric part from the expression and place it along with the variation of the Noether current. Finally, the rest of the terms are identified as the conserved ADT-like current, which can be written as a total derivative of a two-rank anti-symmetric tensor, consisting of the variation of the Noether potential and other identified anti-symmetric terms.  This way of approaching is very convenient, as the Noether current is known for non-Killing cases as well (see the project 8.1 of \cite{paddy}).
 Thereafter, we shall see that our obtained ADT-like current reduces to the standard form of conserved ADT current when $\xi^a$ is considered as a Killing vector.

We now start our analysis by considering the general LL Lagrangian $L(R^{a}_{~bcd}, g^{ab})$. An arbitrary variation of the \la leads to the result
\begin{equation}
\d(\mg L)=\mg E_{ab}\d g^{ab}+\mg\na_a\d v^a~. 
\label{var}
\end{equation}
In the above we denote, $\d v^a=2P^{ibad}(\na_b\d g_{id})$.  Now, as the \la $L$ is invariant under the  \dif $x^a\rightarrow x^a+\xi^a$,   we can obtain the conserved Noether current due to the \dif. Now, under the diffeomorphism, the arbitrary variation $\d$ becomes the Lie variation and from \eqref{var}, one can obtain $\na_a J^a=0$ where, 
\begin{equation}
J^a=2E^{ab}\xi_b+L\xi^a-\lie v^a~, 
\label{JA}
\end{equation}
is the conserved Noether current. Here $\lie$ denotes the Lie derivative of a tensor along the vector $\xi^a$. In the above the crucial step is the first term on the right hand side of (\ref{var}) can be expressed as total derivative term by the generalised Bianchi identity $\nabla_aE^{ab}=0$. Note that in acheving (\ref{JA}), one does not need to use equation of motion at the operational level (for details, see Project $8.1$ in page $394$ of \cite{paddy}).  For some details step see the Appendix B of \cite{Majhi:2014lka}.
Now, this current \eqref{JA} can be further expressed as $J^a=\nabla_bJ^{ab}$ again without using equation of motion, where
\begin{equation}
J^{ab}=2P^{abcd}\na_c\xi_d~,
\label{NJ}
\end{equation}
is the anti-symmetric Noether potential.

Let us now take an arbitrary variation of the above conserved Noether current (\ref{JA}) due to the arbitrary variation of the metric tensor $g_{ab}\rightarrow g_{ab}+h_{ab}$, which leaves the vector $\xi^a$ invariant (\ie, $\d\xi^a=0$, but $\d\xi_a\neq 0$ in general due to the variation in the metric tensor). We then obtain
\begin{align}
\d(\mg J^a)=2\d(\sqrt{-g}E^{ab}\xi_b)-\sqrt{-g}\xi^aE^{ij}h_{ij}
\no 
\\
+\mg (\na_i\d v^i)\xi^a-\d(\mg \lie v^a)~. 
\label{VARJ}
\end{align}
To proceed further, consider the following identity
\begin{equation}
\pounds_{\xi}[\sqrt{-g} \d v^a]=\sqrt{-g}\xi^a\nabla_i\d v^i
-2\sqrt{-g}\nabla_b(\xi^{[a}\d v^{b]})~, 
\label{ID}
\end{equation}
where the notation $A^{[a}B^{b]}=(1/2)(A^aB^b-A^bB^a)$ has been adopted.  Using identity \eqref{ID} in \eqref{VARJ}, it is straightforward to reach
\begin{align}
\d(\mg J^a)-2\sqrt{-g}\nabla_b[\xi^{[a}\d v^{b]}]=2\d(\sqrt{-g}E^{ab}\xi_b)
\no 
\\
-\sqrt{-g}\xi^aE^{ij}h_{ij}-\o^a~, 
\label{VARJ2}
\end{align}
where, 
\begin{equation}
\o^a=\d [\sqrt{-g}\lie v^a]-\pounds_{\xi}[\sqrt{-g} \d v^a]~. 
\label{O1}
\end{equation} 

Next our aim is to collect the terms which can be expressed as covariant derivative of an anti-symmetric tensor and keep them in one side of the equality; while the rest of the terms will be kept on the other side. Note that in Eq. (\ref{VARJ2}), the terms on the left hand side are already in the anti-symmetric form. Let us keep the first two terms on the right hand side unchanged and concentrate on $\o^a$. We shall see that first term of $\o^a$ in (\ref{O1}) has an anti-symmetric part. For that purpose, re-express $\o^a$ as
\begin{eqnarray}
&&\o^a=-\pounds_{\xi}[\sqrt{-g} \d v^a]+2\mg hP^{ibad}\na_b\na_{(d}\xi_{i)}
\no 
\\
&&
+4\mg (\d P^{ibad})\na_b\na_{(d}\xi_{i)} +2\mg P^{ibad}\na_b[\lie h_{id}]
\no 
\\
&&-4\mg P^{ibad}\d \G^l_{bd}\na_{(i}\xi_{l)}~. 
\label{O3}
\end{eqnarray}
The above final expression can be obtained after some involved calculation which is given in the Appendix \ref{App1}.
Here, we have followed the notation $A_{(i}B_{j)}=(1/2)(A_iB_j+A_jB_i)$. Now, identifying the total derivative anti-symmetric part from the last term of $\o^a$ from \eqref{O3}, one can have $\K^a=\na_b\K^{ab}$ where,
\begin{eqnarray}
&&\mg \K^a=\d(\sqrt{-g}E^{ab}\xi_b)-\frac{1}{2}\sqrt{-g}\xi^aE^{ij}h_{ij}
\no
\\
&&+\frac{1}{2}\pounds_{\xi}[\sqrt{-g} \d v^a]-\mg hP^{ibad}\na_b\na_{(d}\xi_{i)}
\no 
\\
&&-2\mg (\d P^{ibad})\na_b\na_{(d}\xi_{i)} -\mg P^{ibad}\na_b[\lie h_{id}]
\no 
\\
&&+\mg P^{ibad}(\na_b h^l_d-\na^lh_{bd})\na_{(i}\xi_{l)}
\no 
\\
&&-\mg P^{ibad}h^l_b\na_d\na_{(i}\xi_{l)}~;
\label{KA}
\end{eqnarray}
and  
\begin{eqnarray}
&&\mg\K^{ab}=\frac{1}{2}\d(\mg J^{ab})-\sqrt{-g}[\xi^{[a}\d v^{b]}]
\no 
\\
&&+\mg P^{abcd} h^l_c\na_{(d}\xi_{l)}~. 
\label{KAKAB}
\end{eqnarray}
Again, we refer our reader to follow the detail calculation from the Appendix \ref{App1}.

Note, that the $\K^{ab}$ is an anti-symmetric tensor. Therefore, the quantity $\K^a$ is conserved for the \dif vector $\xi^a$. In the case of GR, use of equation of motion reduces (\ref{KA}) and (\ref{KAKAB}) to the expression obtained in \cite{Kastor:2002fu} for a $\xi^a$ which was derived following the original approach of ADT \cite{comment1}. On the contrary, the present method is different from this in which Noether prescription has played a central role. We call (\ref{KA}) and (\ref{KAKAB}) as ADT-like conserved current and anti-symmetric potential, respectively. These are the main results of the present paper.  

A point is to be noted here that our current is much robust and more general as we have obtained the conserved current in \ll theory for a wide class of \dif vector, which defines non-Killing (and Killing as well) horizon surfaces. Also, our analysis directly shows the connection between the conserved ADT-like potential and the conserved Noether potential, which is given by (\ref{KAKAB}). Moreover, a careful investigation shows that, in our derivation, we have not directly used the equation of motion $i.e., E^{ab}=0$. In GR, people call this procedure as an ``off-shell'' method. Therefore, in that sense, the obtained conserved ADT-like current and potential is off-shell ones. However, one must be cautious about the fact that the term ``off-shell'' is often misunderstood. The off-shell formalism of obtaining current is just operational in order to obtain the on-shell results (such as the charges). There is another point to consider. As we have not used the equation of motion in our analysis, our results can be used for those null-surfaces as well which might not be obtained from direct solution of the equation of motion.

One must note that, in our analysis, we have repeatedly used $\na_a P^{abcd}=0$. Therefore, the analysis is valid for \ll gravity only. Of course, the same procedure can be extended to other theories of gravity and, in that case, the results must be modified. However, for the time being, we have concentrated only on LL gravity for the following reasons. As we know, one of the key features of GR is that it depends on the geometric properties of the Riemann tensor and the field equations are second order of the dynamical variables. Moreover, GR is ghost free. Now, it is compelling to know whether one can formulate a theory which incorporates higher order polynomials of the curvature tensor, Ricci tensor and Ricci scalar in such a way that the field equations are still second order of the dynamical variables and the theory is free of ghost. The answer is yes! and, the theory is \ll gravity \cite{Lovelock:1971yv, Zwiebach:1985uq, Boulware:1985wk}. Now, while obtaining the conserved current and the potential we do not require to use the equation of motion $E^{ab}=0$ at the operational level but, to compute them explicitly, one has to use a particular background. If one uses the background as a solution of LL gravity, then it will be free of ghosts. However, if the background is not obtained from the LL gravity, one has to carefully check and allow only those backgrounds which does not have any ghost.

Some comments are in order about the above conserved quantities. It is well known that in $U(1)$ Yang-Mills theory due to the gauge symmetry $A_i \longrightarrow A_i + \nabla_i\theta (x)$, one obtains the conserved current as $j^a=\nabla_b[{F^{ab}\theta(x)]}$. When $\theta$ is a constant, it implies that $j^a=0$ because of the equation of motion $\nabla_aF^{ab}=0$. As a result, the conserved charge $Q=\int d^3x j^0=\oint d^2x n_iF^{0i}\theta=0$. The last surface  integral vanishes for a compact surface which encloses the volume. But, it may not vanish for any patch of the surface, which is not a compact one. 

In gravity, when there is a horizon in the spacetime, the whole spacetime is not accessible to the observer. Then the the spacelike volume has two boundaries: one is at the horizon and other one at the infinity. In this case one usually is not interested on the quantity $\int d\Sigma_a\mathcal{K}^a$. Instead the quantities of interest are the surface integral $\int d\Sigma_{ab}\mathcal{K}^{ab}$, evaluated either on the horizon or on the infinity. Both of these two-surfaces are non-compact in nature.  This disconnects (and generalises) the definition of Noether charge from the integral of $\mathcal{K}^a d\Sigma_a$ over some bulk region. Here we call this as the conserved charge. For example, in the case of GR with Killing symmetry, $\int d\Sigma_{ab}\mathcal{K}^{ab}$ leads to ADT conserved quantity calculated on a particular two surface. Of course, in special cases the two integrals are related by Gauss law but the definition which we use in gravity is more general. In particular, it only depends on the integral of $dQ=\mathcal{K}^{ab} d\Sigma_{ab}$ being usefully defined and does not care about the value (zero or non-zero, positive or negative) of $\mathcal{K}^a$ over the bulk.

As mentioned above, we are interested to calculate $\mathcal{K}^{ab}$ over a two surface which is a part of a closed surface, enclosing the spacelike bulk. In this case the diffeomorphism vectors are chosen in a such a way that it defines the surface. For example, if the surface is a generic null surface defined by null vector $l^a$ with the condition $l^al_a=0$, then in that case $\xi^a=l^a$.  In general the expression of $\xi^a$ is obtained by imposing certain condition by which a specific boundary is chosen. 
Moreover, as we have mentioned earlier, even if $\int d\Sigma_a\mathcal{K}^a=0$ in a particular case, $\int d\Sigma_{ab}\mathcal{K}^{ab}$ is not zero in general when it is evaluated on a non-compact two-surface such as the horizon or the asymptotic infinity. Therefore our finding of $\mathcal{K}^{ab}$ in a general way is a very important one for exploring the conserved quantities on a surface which is defined not only through Killing vectors, but also by other types of vectors.

We want to emphasize  once again that our analysis is not only restricted to the Killing horizon, which is defined by the vanishing norm of a Killing vector. As we have mentioned earlier, there are various types of horizons, other than the Killing one. Here we mention couple of them which appear in several discussions and have current interests in different aspects of gravity.
First example is the metric which representing a generic null surface in Gaussian null coordinates (GNC) \cite{Hollands:2006rj, MORALES}:
\begin{equation}
ds^2=-2r\a du^2+2drdu-2r\b_Adx^Adu+\mu_{AB}dx^Adx^B~. 
\label{METRIC}
\end{equation}
Note that the surface is not defined by the vanishing norm of a Killing vector. 
The null horizon in this spacetime (which is at $r=0$) is defined by the vanishing norm of the null vector $l^a=(l^u, l^r, l^{x^A})=(1, 0, \textbf{0})$; and $l_a=(-2r\a, 1, -r\b_A)$. Here, $x^A$ represents all the transverse directions. It can be easily verified that $l^a$ is not a Killing vector (\ie $\pounds_lg_{ab}\neq 0$ in general). Therefore, the existing formalism of the ADT current cannot be used for this null surface.

 Another interesting example is the case of  the scalar--tensor theory. In that (\ie the scalar tensor) theory, a scalar field $\phi$ is non-minimally coupled with the gravity in the original (Jordan) frame described by the metric $g_{ab}$. This coupling can be removed by a conformal transformation $g_{ab}\rightarrow\t g_{ab}=\phi g_{ab}$, along with a rescaling in the scalar field \cite{Bhattacharya:2017pqc, Bhattacharya:2018xlq}. By virtue of these transformations, one arrives to the conformal frame, known as the Einstein frame. If there exists a Killing horizon in the Jordan frame, it is a conformal Killing horizon in the Einstein frame. It is because, a Killing vector (say $\chi$) in Jordan frame, will be a conformal Killing vector in the Einstein frame in general. Mathematically, although $\pounds_{\chi}g_{ab}=0$, we have $\pounds_{\chi}\t g_{ab}=(\pounds_{\chi}\ln\phi)\t g_{ab}$. Now, if $\phi$ evolves with time (\ie $(\pounds_{\chi}\ln\phi)\neq 0$), $\chi$ will no more be Killing vector in Einstein frame and consequently, the horizon in this frame is determined by the conformal Killing vector (not by the Killing one). This is true for any two conformally connected frames. As the existing formalism of ADT current is defined for the Killing vectors, it cannot be used in both of the frames. Apart from these examples, there are a lot of non-Killing horizons in gravity as we have mentioned earlier. Our approach covers all of them (of course for scalar--tensor theory, $E^{ab}$ and $P^{abcd}$ have to be determined accordingly).
 In the following, we shall discuss two cases: one is $\xi^a$ is KV and other one is $\xi^a$ is CKV. 
\section{Special form of diffeomorphism}
Let us now discuss two popular situations. One is the spacetime has Killing symmetry and hence choose $\xi^a$ as Killing vector. Other case is the spacetime has inherent conformal symmetry and so $\xi^a$ can be taken as conformal Killing vector. The former case has been studied extensively in literature while the later one, so far we know, has not been dealt with in this context. 
  
\subsection{$\xi^a$ is a \kv}
If one takes $\xi^a$ as the \kv, satisfying Killing equation $\nabla_a\xi_b+\nabla_b\xi_a=0$, the obtained conserved current $\K^a$ and the potential $\K^{ab}$ (see Eqs. (\ref{KA}) and (\ref{KAKAB})) reduces to the conserved ADT current and the potential \ie
\begin{align}
\mg\K^a|_{KV}=\mg\J^a=\d(\sqrt{-g}E^{ab}\xi_b)
\no 
\\
-\frac{1}{2}\sqrt{-g}\xi^aE^{ij}h_{ij}~,
\label{KillingK1}
\end{align}
and
\begin{align}
\mg\K^{ab}|_{KV}=\mg\J^{ab}
\no
\\
=\frac{1}{2}\d(\mg J^{ab})-\sqrt{-g}\xi^{[a}\d v^{b]}~,
\label{KillingK2}
\end{align}
respectively.
In the above, $\J^a$ is the expression of the usual ADT current.
Exactly the same was obtained earlier in \cite{Kim:2013zha}.
\subsection{$\xi^a$ is a \ckv}
Considering $\xi^a$ as a \ckv~satisfying $\nabla_a\xi_b+\nabla_b\xi_a=(\psi/2)g_{ab}$, and using the identity
\begin{align}
\lie h_{ab}=\frac{\psi}{2}h_{ab}+\frac{\d\psi}{2}g_{ab}~,
\end{align}
which holds only for the \ckv s, the expression (\ref{KA}) reduces to 
\begin{eqnarray}
&&\mg \K^a|_{CKV}=\d(\sqrt{-g}E^{ab}\xi_b)-\frac{1}{2}\sqrt{-g}\xi^aE^{ij}h_{ij}
\no
\\
&&+\frac{1}{2}\pounds_{\xi}[\sqrt{-g} \d v^a]-\frac{\mg}{4}h P_i^{\, \, bai}(\na_b\psi)
\no 
\\
&&-\frac{\mg}{2}g_{id}(\d P^{ibad})(\na_b\psi)-\frac{\mg}{2}P^{ibad}h_{id}(\na_b\psi)
\no 
\\
&&-\frac{\mg}{2}P_i^{\, \, bai}\na_b(\d\psi)~,
\label{KillingK3}
\end{eqnarray}
while (\ref{KAKAB}) yields
\begin{equation}
\mg\K^{ab}|_{CKV}=\frac{1}{2}\d(\mg J^{ab})-\sqrt{-g}\xi^{[a}\d v^{b]}~.
\label{KillingK4}
\end{equation}
Note that although the form of $\K^a|_{CKV}$ is not exactly the same as the $\K^a|_{KV}$ (which is, of course, defined for $\xi^a$ being a \kv), interestingly, the conserved potential $\K^{ab}|_{CKV}$ is related to corresponding Noether potential in exactly identical manner as the conserved ADT current $\K^{ab}|_{KV}$. It is a very new finding which indicates that ADT potential for CKV can be obtained solely by the information of Noether potential and the boundary term in the variation of the action. This is completely similar to KV case.

\subsection{Explicit evaluation of the conserved charges for a black hole spacetime}
Having these conserved quantities, let us now show how to calculate these for a given metric solution of a specific gravitational theory. The charges will be calculated on a non-compact surface. For that we define the charge as
\begin{eqnarray}
\delta Q &=& \frac{1}{16\pi G}\int_{\mathcal{S}}d\Sigma_{ab}\K^{ab}~,
\label{R1}
\end{eqnarray}
where $d\Sigma_{ab}$ is the surface element of $\mathcal{S}$.

The spacetime which we take as an example is the Sultana-Dyer (SD) \cite{Sultana:2005tp} metric which is given as
\begin{equation}
d s^2 =  a^2(t,r)\Big[-F(r)dt^2 + \frac{dr^2}{F(r)}+ r^2(d\theta^2
+\sin^2\theta d\phi^2)\Big]~,
\label{SDSC}
\end{equation}
Here, $F(r) = 1-2M/r$ and the time dependent conformal factor is given as \cite{Faraoni:2013aba, Bhattacharya:2016kbm}:
\begin{equation}
a(t,r) = \Big(t+2M\ln\Big|\frac{r}{2M}-1\Big|\Big)^2~.
\label{a}
\end{equation}
The SD metric is a time-dependent solution of general relativity (GR) in presence of time-like dust and null-like dust (for details see \cite{Sultana:2005tp}). Moreover, as one can see from \eqref{SDSC}, the SD metric is conformal to the Schwarzschild metric. However, unlike the Schwarzschild metric, the SD metric is not asymptotically flat, instead asymptotically it becomes  Friedmann-Lemaitre-Robertson-Walke (FLRW). Moreover, the SD metric has a conformal Killing horizon at $r_{\H}=2M$ \cite{Majhi:2014lka} and the conformal Killing horizon is located where the conformal Killing vector $\xi^a$ of SD background becomes null \ie $\xi^a\xi_a=0$. Note, that the \ckv~of SD background is given by $\xi^a=(1,0,0,0)$ and $\xi_a=a^2(t,r)(-F(r),0,0,0)$. Here we shall evaluate (\ref{R1}) at the two different surfaces: at the horizon and at the assymptotic region which is FLRW. Since the spacetime (\ref{SDSC}) has inherent conformal Killing vector, the expression for $\K_{ab}$ must be given by (\ref{KillingK4}). In this case (\ref{R1}) can be re-expressed as 
\begin{align}
\d Q=\frac{1}{32\pi }\d\int_{\mathcal{S}}d\Sigma_{ab} J^{ab}
-\frac{1}{16\pi }\int_{\mathcal{S}}d\Sigma_{ab}(\xi^{[a}\d v^{b]})~. \label{R2}
\end{align}
In the above the first term of the right hand side is obtained in the following way. Originally, it is given by $\int d\Sigma_{ab}\delta(\sqrt{-g}J^{ab})/\sqrt{-g}$. Now the surface element is constructed as $d\Sigma_{ab}=d^2x\sqrt{\sigma}(t_ar_b-t_br_a)$ where $t_a$ and $r_a$ are chosen either in spacelike-timelike basis or in null-null basis, whereas $\sigma$ is the determinant of the induced metric on $\mathcal{S}$. This element is also be expressed as $d\Sigma_{ab}=d^2x\sqrt{-g}(\hat{t}_a\hat{r}_b-\hat{t}_b\hat{r}_a)$ where the hated vectors just determine the direction of the actual vectors and are constant. Then, since $\delta(\hat{t}_a\hat{r}_b-\hat{t}_b\hat{r}_a)=0$, the integration is transformed to the first term of (\ref{R2}).

First let us calculate (\ref{R2}) on the conformal Killing horizon. In this case we choose $r^a$ as the conformal Killing vector $\xi^a$ of the spcaetime and $t^a$ is constructed to be the auxilary null vector $l^a$, such that $l.\xi=-1$. Then one finds that
no contribution comes from the second integral of \eqref{R2} as $\xi^2$ and $\xi_a$ both vanish on the horizon. 
The first term of (\ref{R2}) is evaluated in the following way.
Using the definition of \eqref{NJ}, for GR one obtains $J^{ab}=\na^a\xi^b-\na^b\xi^a$ which by conformal Killing equation $\nabla_a\xi_b+\nabla_b\xi_b=\Omega^2g_{ab}$ yields $J^{ab} = \Omega^2g_{ab}-2\nabla_b\xi_a$ with $\Omega^2 = \nabla_a\xi^a/2$. In presence of conformal Killing vector, one can define a surface gravity $\kappa$ as $\nabla_a\xi^2=-2\kappa\xi_a$ which is constant on the horizon \cite{RR1}. Using these, the first integration is evaluated to be as
\begin{eqnarray}
&&\frac{1}{32\pi }\int_{\H}d\Sigma_{ab}J^{ab} 
\nonumber
\\
&&= -\frac{1}{16\pi }\int_{\H}d^2x\sqrt{\sigma}\xi_al_b(-2\nabla^b\xi^a+\Omega^2g^{ab})
\nonumber
\\
&&= \frac{1}{16\pi }\int_{\H}d^2x\sqrt{\sigma}(l^a\nabla_a\xi^2+\Omega^2)
\nonumber
\\
&&= \frac{1}{16\pi }\int_{\H}d^2x\sqrt{\sigma}(2\kappa+\Omega^2)~,
\end{eqnarray}
where in the last step $\nabla_a\xi^2=-2\kappa\xi_a$ and $\xi\cdot l=-1$ have been used. For SD metric, one finds $\Omega^2=4/\sqrt{a}$ and then 
\begin{equation}
\d Q_{\H}=\frac{1}{16\pi }\int_{\H}d^2x\sqrt{\sigma}(2\kappa+\frac{4}{\sqrt{a}})~. \label{delqh}
\end{equation}
Since $\kappa$ is a constant and $a(t,r)$ is independent of the angular coordinates, whereas the integration is performed on the angular coordinates, \eqref{delqh} implies 
\begin{equation}
\d Q_{\H}=\frac{1}{2}r_{\H}^2a_{\H}^2(\kappa+\frac{2}{\sqrt{a_{\H}}})~.
\end{equation}
Further as $a_{\H}$ diverges, we consider only the leading order contribution which provides
\begin{align}
\d Q_{\H}=\d (\frac{\kappa}{2\pi }\pi a_{\H}^2r_{\H}^2)~. 
\end{align}
Although, here we have obtained the charge at the horizon, we cannot concretely say what physical quantity the charge implies. However, from the known result of \cite{Majhi:2014hpa}, we can identify $\pi a_{\H}^2r_{\H}^2$ as the entropy of the SD black hole. Thus, we can say that entropy is given in terms of our defined charge as $S=\frac{2\pi}{\kappa}Q_{\H}$.

Now, we shall calculate the conserved charge in the asymptotic region on a $r=$constant and $t=$constant surface. We denote this asymptotic two-surface as $\partial c$.
In this case, we write the elemental surface area of two-surface $d\Sigma_{ab}$ in terms of timelike and spacelike normals as  $d\Sigma_{ab}=(n_as_b-n_b s_a)\sqrt{\sigma}d^2x$, with $n^a n_a=-1$ and $ s^as_a=1$~. The normal are found to be 
\begin{equation}
n^a=(\frac{1}{a\sqrt{F}}, 0, 0, 0); \,\,\ s^a=(0, \frac{\sqrt{F}}{a}, 0, 0)~.
\label{R3}
\end{equation}
With this, the first integral of \eqref{R2} yields 
\begin{equation}
\frac{1}{32\pi}\d\int_{\p c}d\Sigma_{ab} J^{ab}=\frac{1}{2}[M a^2+4 Mra\sqrt{a}]~.
\label{R4}
\end{equation} 
The second integral of \eqref{R2} provides $-\frac{1}{16\pi}\int_{\p c}d\Sigma_{ab}(\xi^{[a}\d v^{b]})=\frac{1}{16\pi}\int_{\p c}\sqrt{ \sigma}a^2\d  v^r$. From the definition of $\d v^a$, we further obtain in this case $\d v^r=- g^{rr} g^{tt}\na_r\d g_{tt}$. Now, to compute $\d g_{tt}$, one has remember that the SD spacetime is asymptotically FLRW where the metric component $ g_{tt}\rightarrow  g_{tt}=-a^2$~. Therefore, $\d  g_{tt}=\frac{2Ma^2}{r}$. Using this, finally one can obtain $\d v^r=\frac{1}{a^4}[\frac{2Ma^2}{r^2}-\frac{4Ma^2}{r^3F}-\frac{32M^2a^{\frac{3}{2}}}{r^2F}]$. Then the second integration yields
\begin{equation} 
-\frac{1}{16\pi}\int_{\p c}d\Sigma_{ab}(\xi^{[a}\d v^{b]})=\frac{1}{2}[Ma^2-\frac{2Ma^2}{rF}-\frac{16M^2a^{\frac{3}{2}}}{F}]~.
\end{equation} 
Therefore, finally we obtain 
\begin{align}
\d Q_{\p c}=\Big[Ma^2+2 Mra^{\frac{3}{2}}-\frac{Ma^2}{rF}-\frac{8M^2a^{\frac{3}{2}}}{F}\Big]~.
\end{align}
Here, we have shown the procedure how to compute those charges in the two region. However, for the present moment, we cannot identify (using any first principal) what macroscopic quantities of a black hole (like mass, angular momentum, entropy, etc) those charge imply. In the following section, we shall discuss that if someone tries to obtain the first law following the ADT approach for the conformal Killing horizon with the charges and potential that we have obtained, one has to deal with several non-trivial terms. Since the original ADT approach is based on the Killing symmetry, those term vanishes. Presently, we cannot say what the extra terms, which appears for the \ckv s, contributes to. Therefore, using the original formalism, we cannot identify mass, angular momentum \etc~for the conformal Killing horizon. This topic is now under investigation.

\section{Summary and outlook}
Defining thermodynamical entities and establishing the thermodynamic laws for a non-Killing dynamical horizon has been a major challenge in gravity. Several physicists (including us) are trying to obtain the thermodynamic first law for a generic null horizon or for a conformal Killing horizon. What has been understood so far is that one has to start developing everything from the very basics. As it is well-known, the root of the black hole thermodynamics lies in the conserved current in the theory because, all the physical thermodynamic parameters can be obtained from the conserved current. Therefore, it is very much needed to generalize the conserved quantities for a non-Killing symmetry.

It appears that the ADT approach is a very successful method to obtain the thermodynamic description from the conserved currents without any ambiguity like anomalous two factor in Komar case \cite{Komar:1958wp}. But till now all attempt has been done for existence of a Killing vector in the spacetime. Since in reality we need to encounter non-Killing situation, it is now necessary to obtain the conserved quantities for any diffeomorphism vector. So far it has been successfully done for Noether conserved quantities.  In this letter, we have formulated an elegant approach to define an ADT-like current for a horizon defining \dif vector in \ll gravity as the original ADT approach is valid only for the \kv s. 

Here, we have taken the general LL Lagrangian. Then we have varied the Lagrangian to obtain the Noether current due to the diffeomorphism. Thereafter, our purpose was to obtain the ADT-like current for non-Killing \dif vector. We have defined the conserved current as the derivative of a two-rank anti-symmetric conserved potential. Then we have taken two limits: $\xi^a$ as a \kv, and then $\xi^a$ as a \ckv. For \kv~case, the entire result reduces to the original ADT currents and the potential, defined for the \kv s. We obtain even more interesting result for the \ckv s. In this case, the expression of conserved potential is exactly similar to the ADT potential. However, the expression of the conserved current differs. Following our method, one can show the connection of the conserved potential with the conserved Noether potential in each case.

As we have emphasized several times, the entire analysis is very general in every aspects. The currents and the potentials are defined for wide class of \dif vector, which defines horizon surfaces. Also, the analysis has been made for the general \ll gravity form which one can obtain the corresponding results for the GR or for any other higher order gravity, and yet nowhere the equation of motion has been used. Thus, we believe, our result will be an useful one in the black hole thermodynamics.

Before concluding, we want to mention an important future aspect of our analysis. Till now we are familiar with the Wald's approach \cite{Wald:1993nt} to find the first law of thermodynamics for a Killing vector. Now the question is: Can we follow the similar prescription to find a reasonable form of first law in the case for a null surface which is not accompanied by a Killing symmetry. Probably, our most general relation $\mathcal{K}^a=\nabla_b\mathcal{\K}^{ab}$ with the quantities are given by equations (\ref{KA}) and (\ref{KAKAB}), inherently captures that general structure. To get a feel of the generalization, let us first consider the Killing situation. In this case $\mathcal{K}^a$ and $\mathcal{K}^{ab}$ are given by (\ref{KillingK1}) and (\ref{KillingK2}), respectively. On-shell (i.e. using equation of motion) $\mathcal{K}^a$ vanishes and then using the steps of Wald, one obtains the first law of black hole mechanics. This is all known. But, for a conformal Killing vector the situation is a bit different. For simplicity, we consider GR theory with conformal symmetry.   
Here $P^{abcd}=(1/2)(g^{ac}g^{bd}-g^{ad}g^{bc})$, and then for on-shell condition, the relation $\mathcal{K}^a=\nabla_b\mathcal{\K}^{ab}$ reduces to the following form
\begin{align}
\frac{1}{2}\d(\mg J^{a})-\sqrt{-g}\na_b[\xi^{[a}\d v^{b]}]
\no
\\
=\frac{\mg}{4}\Big[\psi\na_ih^{ai}+h(\na^a\psi)-\psi(\na^ah)-h^{ai}\na_i\psi\Big]~, 
\label{CKVGR}
\end{align}
where, $\psi=\na_i\xi^i$. This clearly indicates that the terms on the right hand side of the above equation contributes to thermodynamic quantities in this case. Note that they vanishes for the Killing situation. Hence it is evident that for most general case the usual definition of thermodynamic quantities (like entropy is given by the Noether charge calculated on the horizon) can not be taken as general one. Therefore it would be interesting to see how the terms like those appear on the right hand side of (\ref{CKVGR}) modify the definition of several thermodynamic quantities. Till now we do not have any conclusive statement. The investigations are going on in this direction. Hope we shall be able to report soon.
\section*{Appendix}
\appendix
\section{\label{App1}Derivation of the Eqs. (\ref{O3}), (\ref{KA}) and (\ref{KAKAB})}
We know that
\begin{eqnarray}
\lie v^a=4p^{ibad}\na_b\na_{(i}\xi_{d)}~. \label{A1}
\end{eqnarray}
Thus,
\begin{eqnarray}
&&\d [\sqrt{-g}\lie v^a]=2\mg hP^{ibad}\na_b\na_{(i}\xi_{d)}
\no 
\\
&&+4\mg (\d P^{ibad})\na_b\na_{(i}\xi_{d)} +4\mg  P^{ibad}\d\{\na_b\na_{(i}\xi_{d)} \}~.
\no 
\\ \label{A2}
\end{eqnarray}
Now, a detail calculation shows that 
\begin{align}
4P^{ibad}\d\{\na_b\na_{(i}\xi_{d)} \}=2P^{ibad}\na_b[\lie h_{id}]
\no 
\\
-4P^{ibad}\d \G^l_{bd}\na_{(i}\xi_{l)}. \label{A3}
\end{align}
Replacing \eqref{A2} and \eqref{A3} in \eqref{O1} one can obtain \eqref{O3}.

Now, 
\begin{align}
4P^{ibad}\d \G^l_{bd}\na_{(i}\xi_{l)}=2P^{ibad}[\na_b h^l_d-\na^lh_{bd}]\na_{(i}\xi_{l)}
\no 
\\
+2\na_b[P^{abic} h^l_c\na_{(i}\xi_{l)}]-2P^{ibad}h^l_b\na_d\na_{(i}\xi_{l)}
\end{align}
Thus, from \eqref{VARJ2} one can obtain
\begin{eqnarray}
&&\d(\mg J^a)-2\sqrt{-g}\nabla_b[\xi^{[a}\d v^{b]}]=2\d(\sqrt{-g}E^{ab}\xi_b)
\no 
\\
&&-\sqrt{-g}\xi^aE^{ij}h_{ij}+\pounds_{\xi}[\sqrt{-g} \d v^a]-2\mg hP^{ibad}\na_b\na_{(d}\xi_{i)}
\no 
\\
&&-4\mg (\d P^{ibad})\na_b\na_{(d}\xi_{i)} -2\mg P^{ibad}\na_b[\lie h_{id}]
\no 
\\
&&+2\mg P^{ibad}[\na_b h^l_d-\na^lh_{bd}]\na_{(i}\xi_{l)}
\no 
\\
&&-2\mg P^{ibad}h^l_b\na_d\na_{(i}\xi_{l)}+2\mg\na_b[P^{abic} h^l_c\na_{(i}\xi_{l)}]~.
\no \\
\end{eqnarray}
Replacing the last term on the left hand side, one can obtain the desired results given in \eqref{KA} and \eqref{KAKAB}.

\end{document}